\documentclass[seceq,preprint]{ptptex}
\usepackage{graphicx}
\usepackage{wrapft}

\title{
Statistical properties of spectral fluctuations for a quantum system with infinitely many components}

\author{
H.~Makino$^1$, N.~Minami$^2$, and S.~Tasaki$^3$
}

\inst{
$^1$Department of Human and Information Science, Tokai University, 1117 Kitakaname, 
Hiratsuka-shi, Kanagawa 259-1292, Japan\\
$^2$School of Medicine, Keio University, Hiyoshi 4-1-1, Kohoku-ku, Yokohama, Kanagawa 
223-8521, Japan\\
$^3$Department of Applied Physics, Waseda University, 3-4-1 Okubo, Shinjuku-ku, Tokyo 
169-8555, Japan
}

\recdate{
Jan 1, 2009}

\abst{
Extending the idea formulated in Makino {\it{et al}}[Phys.Rev.E {\bf{67}},066205],
that is based on the Berry--Robnik approach [M.V. Berry and M. Robnik, J. Phys. A {\bf{17}}, 2413], we 
investigate the statistical properties of a two-point spectral correlation for a classically 
integrable quantum system.  The eigenenergy sequence of this system is regarded as a 
superposition of infinitely many independent components in the semiclassical limit.  We derive the level 
number variance (LNV) in the limit of infinitely many components and discuss its deviations from Poisson statistics. The slope of the limiting LNV is found to be larger than that of Poisson statistics when the individual components have a certain accumulation.  This property agrees with the result from the semiclassical periodic-orbit theory that is applied to a system with degenerate torus actions[D. Biswas, M.Azam,and S.V.Lawande, Phys. Rev. A {\bf 43}, 5694].
}

\begin{document}
\maketitle

\section{Introduction}
\label{sect1}
Study of energy level statistics has played an important role in elucidating the universal properties of quantum systems, which in the semiclassical limit, reflect regular and chaotic features of classical dynamics.  Berry and Tabor conjectured that the eigenenergy levels of a quantum system, whose classical dynamical system is integrable, have the same fluctuation property as uncorrelated random numbers from a Poisson process, and thus are characterized by Poisson statistics\cite{BT77}.  This conjecture is in contrast with the conjecture of Bohigas, Giannoni, and Schmit, which assert that GOE or GUE statistics based on random matrix theory (RMT) are applicable to the fluctuation property of energy levels of a quantum system whose classical dynamical system is fully chaotic\cite{BGS84}.  These contrasting conjectures have been examined using various statistical observables, e.g., the nearest-neighbor level-spacing distribution (NNLSD) and the level number variance (LNV)\cite{ME91}.\\
\indent
The NNLSD $P(S)$ is the observable most commonly used to study short-range fluctuations in a spectrum.  For eigenenergy levels on the unfolded scale\cite{Bo89}, this observable is defined as the probability density of finding a distance $S$ between adjacent levels.  In Poisson statistics, it is characterized by the exponential distribution $P(S)=e^{-S}$; on the other hand, in GOE/GUE statistics, it approximates the Wigner distribution very well\cite{ME91}.\\
\indent
The LNV $\Sigma^2(L)$ is the observable commonly used to study correlations between pairs of levels, which characterizes long-range 
fluctuations in a spectrum.  It is defined as the average variance of the number of 
levels in an energy interval containing an average of $L$ levels.  On the unfolded scale, this interval is equivalent to an interval of length $L$, and the LNV is thus defined as
\begin{equation}
\Sigma^2(L)=\langle\left({\mathcal{N}}(\epsilon+L)-{\mathcal{N}}(\epsilon)-L\right)^2\rangle, \label{EQ.(1.1)}
\end{equation}
where ${\mathcal{N}}(\epsilon)$ is the number of eigenenergy levels below $\epsilon$, and the brackets $\langle\cdots\rangle$ represent the average over the value of $\epsilon$.  In Poisson statistics, the LNV is equal to the number itself $\left[\mbox{i.e., }\Sigma^2_{\mbox{\tiny Poisson}}(L)=L\right]$; on the other hand, in RMT, in which $\beta=1$ and $2$ correspond to the GOE and GUE level statistics, respectively, the LNV increases logarithmically with $L$ $\left[\mbox{i.e., } \Sigma^2_{\beta}(L)\sim(2/\beta\pi^2)\ln(2\pi L) \right]$\cite{ME91}.  According to Berry's semiclassical theory\cite{Be85}, the LNV of a quantum system with $f$ degrees-of-freedom should display these universal statistics in the region $L<<L_{\mbox{\tiny max}}\sim \hbar^{-(f-1)}$ as a consequence of the uniform distributed property of periodic orbits in the phase space\cite{HO84}.  In the semiclassical limit, in which the Planck constant tends to zero $(\hbar\to 0)$, one expects to observe the universalities for all $L\geq0$ in systems with $f \geq 2$.\\
\indent
There are many works examining the Berry-Tabor conjecture in classically integrable 
quantum systems\cite{Be85,BT76,Mo81,Bl90,Sa97,CK97,Ma98,Ma01,Es,CCG85,Fe85,SV86,RV98}.  
Although the mechanism that supports this conjecture remains to be clarified; the 
statistical property of eigenenergy levels to be characterized by Poisson 
statistics is now widely accepted as a universal property of generic integrable 
quantum systems in the semiclassical limit.\\
\indent
One possible mechanism producing Poisson statistics has been proposed by 
Makino {\it{et al}}\cite{MT03}, based on the Berry-Robnik approach\cite{BR84}.  
We briefly review the outline below: For an integrable 
system, individual orbits are confined in each inherent torus whose surface 
is defined by holding its action variable constant, and the whole region of the phase space is densely covered with invariant tori.  In other words, the phase space consists of infinitely many regions, which have infinitesimal volumes in Liouville measure.  Because of the suppression of quantum tunneling in the semiclassical limit, the Wigner functions of quantal eigenstates are expected to be localized in the phase space region explored by a typical trajectory\cite{Be77,Ro98}, and to form independent components.  For a classically integrable quantum system, 
the Wigner function localizes on the infinitesimal region in $\hbar\to0$ 
and tends to a $\delta$ function on a torus that is designated 
by quantum numbers\cite{Be1977}.  Then, the eigenenergy levels can be 
represented as a statistically independent superposition of infinitely 
many components, each of which gives an infinitesimal contribution to 
the level statistics.  Therefore, if the individual spectral components 
are sparse enough, one would expect Poisson statistics to apply here, 
as a result of the law of small numbers\cite{Fe57}.\\
\indent
The statistical independence of spectral components is assumed to be justified by the principle of uniform semiclassical condensation of eigenstates in the phase space and by the lack of their mutual overlap, and thus, can only be expected in the semiclassical limit\cite{Ro98,LR94}.  This mechanism was introduced as a basis for the Berry-Robnik approach to investigate the relation between the phase space geometry and the formation of energy level sequence in a generic mixed quantum system, whose classical dynamics is partly regular and partly chaotic\cite{BR84}.  The validity of the Berry-Robnik approach is confirmed by numerical computations for the mixed quantum systems in the extremely deep semiclassical region\cite{PR93,Pr96}, which is called the {\it{Berry-Robnik regime}}\cite{PR97}.\\
\indent 
Based on this view, Makino and Tasaki investigated the short-range spectral statistics of classically integrable systems\cite{MT03}.  They derived the cumulative function of NNLSD, i.e., $M(S)=\int_0^S P(x) dx$, is derived in the limit of infinitely many components, which is characterized by a single monotonically increasing function $\bar{\mu}(0;S)$ of the nearest level spacing $S$ as 
\begin{equation}
M_{\bar{\mu}}(S)=1-\left[1-\bar{\mu}(0;S)\right]
\exp{\left(-\int_0^S\left[ 1-\bar{\mu}(0;x)\right]dx\right)}.
\label{EQ.(1.2)}
\end{equation}
The function $\bar{\mu}(0;S)$ classifies $M_{\bar{\mu}}(S)$ into three cases: Case 1, Poisson distribution $M_{\bar{\mu}}(S)=1-e^{-S}$ for ${}^\forall S\geq 0$ if $\bar{\mu}(0;+\infty)=0$; Case 2, asymptotic Poisson distribution, which converges to the Poisson distribution for $S\to+\infty$, but possibly not for small spacings $S$ if $0<\bar{\mu}(0;+\infty)<1$; and Case 3, sub-Poisson distribution, which converges to $1$ for $S\to+\infty$ more slowly than does the Poisson distribution, if $\bar{\mu}(0;+\infty)=1$.  Therefore, the Berry-Robnik approach when applied to classically integrable quantum systems, admits deviations from Poisson statistics.\\
\indent 
Cases 2 and 3 are possible when the eigenenergy levels of individual components show a singular NNLSD result from strong accumulation\cite{MT03} (see also Section {\ref{sect3}} of the present paper), which is expected when the classical dynamical systems have a spatial symmetry\cite{BT77,CK97,Es,SV86,RV98,RB81,BW84,Sh89,BAL91} 
or a time-reversal symmetry\cite{Sh75,CS95,FS97}.  One possible example is a rectangular billiard with a rational ratio of squared sides\cite{BT77,CK97,RV98,Sh89,BAL91}.  These results suggest the existence of a new statistical 
law different from Poisson statistics in the strongly-degenerate quantum systems, and raises the question as to whether 
similar behaviors appear in the two-point spectral correlation.\\
\indent
Deviation from Poisson statistics, resulting from the symmetry, is also exhibited within the framework of the periodic-orbit theory\cite{Gu70}.  Based on the semiclassical theory of Berry and Tabor\cite{BT76}, Biswas, Azam, and 
Lawande investigated the LNV for classically integrable systems with degeneracy in orbit actions\cite{BAL91}.  They 
showed that the slope $g_{\mbox{\tiny{av}}}(L)$ of the LNV $\Sigma^2(L)=g_{\mbox{\tiny{av}}}(L) L$ is described by the average degeneracy of actions of the periodic orbits, which for the degenerated quantum systems, is greater than $1$, and it is the slope of the variance of Poisson statistics.  This result leads to the possibility that the degeneracy of actions, induced by the symmetry, could be an essential factor for the singularity of individual spectral components that yields Cases 2 and 3.  This possibility is confirmed by examining the properties of LNV for Cases 2 and 3.\\
\indent
In this study, extending the theory of Makino and Tasaki\cite{MT03}, we investigate the LNV $\Sigma^2(L)$ of quantum systems whose energy levels consist of infinitely many independent components, and show that the non-Poisson limits (Cases 2 and 3) are possibly observed also in the long-range spectral fluctuations.\\
\indent
Based on the Berry-Robnik approach, the overall LNV is derived as follows: We consider a system whose classical phase space is decomposed into $N$ disjoint regions that give the distinct spectral components.  The Liouville measures of these regions are denoted by $\rho_n (n=1,2,3,\cdots,N)$, which satisfy $\sum_{n=1}^N\rho_n=1$.  Let $E(k;L),k=0,1,2,\cdots$ be the distribution function, which denotes the probability to find $k$ levels in an interval $(0,L)$\cite{ME91,MC72,ABS97}.   
The LNV $\Sigma^2(L)$ is expressed by $E(k;L)$ as
\begin{equation}
\Sigma^2(L) = \sum_{k=0}^{+\infty}(k-L)^2 E(k;L).
\label{EQ.(1.3)}
\end{equation}
Let $P(k;S),k=0,1,2,\cdots,$ be the level-spacing distribution, which denotes the probability density to find $k$ levels in an interval of length $S$ beginning at an arbitrary level $\epsilon_i$ and ending at the level $\epsilon_{i+k+1}$.  $P(k;S)$ is related to $E(k;L)$ as
\begin{equation}
P(k;L) = \frac{\partial^2}{\partial L^2}\sum_{j=0}^k (k-j+1) E(j;L),
\label{EQ.(1.5)}
\end{equation}
and 
\begin{equation}
E(k;L) = \int_L^{+\infty}dx 
\int_x^{+\infty}\left[ P(k;S)-2P(k-1;S)+P(k-2;S)\right]dS,
\label{EQ.(1.4)}
\end{equation}
where $P(j<0;S)=0$\cite{ME91,MC72,ABS97}.
Eqs.(\ref{EQ.(1.5)}) and (\ref{EQ.(1.4)}) are known as the formulae in the theory of point process, and are derived as corollaries of the Palm-Khinchin formula\cite{MI07}.
The NNLSD, which was expressed by $P(S)$ at the beginning of this section, is the special case with $k=0$.\\
\indent
When the entire sequence of energy levels is a product of statistically independent superposition of $N$ subsequences, $E(k;L)$ is decomposed into those of subsequences, $e_n(k;L)$ as\cite{ME91,Po60}
\begin{equation}
E_N(k;L)=\sum_{\sum_{n=1}^N k_n=k}\prod_{n=1}^N e_n(k_n;\rho_n L),
\label{EQ.(1.6)}
\end{equation}
where $e_n$ satisfies the normalization conditions
\begin{equation}
\sum_{k=0}^{+\infty}e_n(k;\rho_n L)=1
\label{EQ.(1.7)}
\end{equation}
and
\begin{equation}
\sum_{k=0}^{+\infty}k e_n(k;\rho_n L)=\rho_n L.
\label{EQ.(1.8)}
\end{equation}
In terms of normalized level-spacing distribution $p_n(k;S)$ of the subsequence, $e_n(k;L)$ is specified as
\begin{equation}
e_n(k;L)=\rho_n\int_L^{+\infty}dx\int_x^{+\infty}\left[ p_n(k;S)-2p_n(k-1;S)+p_n(k-2;S) \right]dS,
\label{EQ.(1.9)}
\end{equation}
where $p_n(j<0;S)=0$, and $p_n(k;S)$ satisfies the normalizations
\begin{equation}
\int_0^{+\infty}p_n(k;S)dS=1
\label{EQ.(1.10)}
\end{equation}
and
\begin{equation}
\int_0^{+\infty}S p_n(k;S)dS=\frac{k+1}{\rho_n}.
\label{EQ.(1.11)}
\end{equation}
Note that in general, the individual components 
are not always unfolded automatically even when the overall spectrum is unfolded.  However, in 
a sufficiently small interval $[\epsilon,\epsilon+\Delta\epsilon ]$, each spectral component obeys the same scaling law (see Appendix of Makino {\it{et al}}\cite{MT03}) and thus is unfolded automatically by an overall unfolding procedure.  In the Berry-Robnik approach, Eq.(\ref{EQ.(1.6)}) relates the level statistics in the semiclassical limit with the phase space geometry.\\
\indent
In most general cases, the individual components might have degeneracy of levels that lead to singular level spacing distributions.  In such a case, it is convenient to use its cumulative distribution function $\mu_n(k;S)$,
\begin{equation}
\mu_n(k;S)=\int_0^S p_n(k;x)dx.
\label{EQ.(1.12)}
\end{equation}
This function satisfies
\begin{equation}
\mu_n(k-1;S)\geq \mu_n(k;S)\quad\mbox{for all}\quad k\geq 1\ \mbox{and for all}\quad S\geq0,
\label{EQ.(1.13)}
\end{equation}
and the normalization condition
\begin{equation}
\sum_{k=0}^{+\infty}\left[\mu_n(k-1;S)-\mu_n(k;S)\right] = 0,
\label{EQ.(1.14)}
\end{equation}
with $\mu_n(j=-1;S)=0$, since $f_n(k;S)$,
\begin{equation}
f_n(k;S)= \left\{
\begin{array}{lc}
\mu_n(k-1;S)-\mu_n(k;S),&\qquad k\geq1\nonumber\\
1-\mu_n(0;S),&\qquad k=0\nonumber
\end{array}\right.\label{EQ.(1.15)}
\end{equation}
denotes the probability to find $k$ levels in an interval of length $S$ beginning at an arbitrary level, and obviously satisfies, from the definition of this function, the conditions $f_n(k;S)\geq0$ and $\sum_{k=0}^{+\infty}f_n(k;S)=1$.\\
\indent
In addition to Eq.(\ref{EQ.(1.6)}), we introduce the following two assumptions that were introduced in Ref.\cite{MT03}:\\
\indent
{\it{Assumption}} (i). The statistical weights of independent regions vanish uniformly in the limit of infinitely many regions:
\begin{equation}
\max_n \rho_n \rightarrow 0\quad \mbox{as}\quad N\rightarrow +\infty.
\label{EQ.(1.16)}
\end{equation}
\indent
{\it{Assumption}} (ii). For $k=0,1,2,\cdots$, the weighted mean of the cumulative distribution of energy spacing, namely,
\begin{equation}
\mu(k;x)=\sum_{n=1}^N \rho_n\mu_n(k;x)
\label{EQ.(1.17)}
\end{equation}
converges as $N\to +\infty$ to $\bar{\mu}(k;x)$,
\begin{equation}
\lim_{N\to +\infty}\mu(k;x) = \bar{\mu}(k;x),
\label{EQ.(1.18)}
\end{equation}
where the convergence is uniform on each closed interval: $0\leq x \leq S$.  In the Berry-Robnik approach, the statistical weights of individual components are the phase volumes (Liouville measures) of the corresponding invariant regions.\\
\indent
Under the Assumptions (i) and (ii), Eqs.(\ref{EQ.(1.3)}) and (\ref{EQ.(1.6)}), we 
obtain the overall LNV in the limit $N\to +\infty$:
\begin{equation}
\Sigma_{\bar{\mu}}^2(L)= L +2 \int_0^L \sum_{k=0}^{+\infty} \bar{\mu}(k;S)dS.
\label{EQ.(1.19)}
\end{equation}
\indent
When $\bar{\mu}(k;S)=0$ for all $k$, the LNV of the whole energy sequence reduces to $\Sigma^2_{\mbox{\tiny Poisson}}(L)=L$.  This condition is expected when 
the individual components are sufficiently sparse.  In general, we expect 
$\bar{\mu}(k;S) > 0$, which corresponds to a certain accumulation of the levels of individual components.  In this case, the LNV of the whole energy sequence deviates from $\Sigma^2_{\mbox{\tiny Poisson}}(L)$ in such a way that the slope of the variance is greater than $1$. \\
\indent
The present paper is organized as follows. The limiting LNV (\ref{EQ.(1.19)}) is derived from Eqs.(\ref{EQ.(1.3)}) and (\ref{EQ.(1.6)}), and Assumptions (i) and (ii) in Section \ref{sect2}.  In Section \ref{sect3}, the property of the limiting LNV is analyzed for the Cases 1--3, wherein the deviations from Poisson statistics are clearly observed in Cases 2 and 3.  We present a numerical analysis for the rectangular billiard that shows deviations from Poisson statistics in Section \ref{sect4}.  In the concluding section, we discuss some relationships between our results and other related works.
%
\section{limiting level number variance}
\label{sect2}
In this section, by using Eqs.(\ref{EQ.(1.6)}) and (\ref{EQ.(1.3)}), and Assumptions (i) and (ii) introduced in Section\ref{sect1}, we derive the limiting LNV,
\begin{equation}
\Sigma^2_{\bar{\mu}}(L)=L+2\int_0^L \sum_{k=0}^{+\infty}\bar{\mu}(k;S)dS,
\label{EQ.(2.1)}
\end{equation}
in the limit of infinitely many components $N\to+\infty$.\\
\indent
First, we rewrite Eq.(\ref{EQ.(1.3)}) in terms of the function $e_n$, and decompose it into the LNV $\sigma_n^2$ of individual components:
\begin{eqnarray}
\Sigma_N^2(L)&=&\sum_{k=0}^{+\infty}(k-L)^2 \sum_{\sum_{n=1}^N k_n=k}\prod_{n=1}^N e_n(k_n;\rho_n L)\label{EQ.(2.2)}\\
&=& \sum_{n=1}^N \sum_{k=0}^{+\infty}(k-\rho_n L)^2 e_n(k;\rho_n L) \equiv \sum_{n=1}^N \sigma_n^2 
(\rho_n L),\label{EQ.(2.3)}
\end{eqnarray}
where we have used the properties $\sum_{k=0}^{+\infty}E_N(k;L)=1$, $\sum_{k=0}^{+\infty}k E_N (k;L)=L$, and 
\begin{equation}
\sum_{k=0}^{+\infty}k^2 E_N(k;L)= \sum_{n=1}^N \sum_{k_n=0}^{+\infty}k_n^2 e_n(k_n;\rho_n L) +  L^2-\sum_{n=1}^N(\rho_n L)^2,
\label{EQ.(2.4)}
\end{equation}
which follow from Eqs.(\ref{EQ.(1.7)}) and (\ref{EQ.(1.8)}), and the relationships $k=\sum_{n=1}^N k_n$ and $\sum_{n=1}^N\rho_n =1$ (see also Appendix A). 
The functions $e_n(k;\rho_n L)$ in the above equation are rewritten in terms of the cumulative level-spacing distribution functions $\mu_n(k;S)$ of individual components.
\begin{eqnarray}
e_n(k;\rho_n L) 
&=&\left\{
\begin{array}{lc}
-\rho_n \int_L^{+\infty} dS \left[\mu_n(k;S)-2\mu_n(k-1;S)+\mu_n(k-2;S)\right],&\qquad k\geq 2 \\
-\rho_n \int_L^{+\infty} dS \left[\mu_n(1;S)-2\mu_n(0;S)+1\right],&\qquad k=1\\
\rho_n \int_L^{+\infty} dS \left[1-\mu_n(0;S) \right],&\qquad k=0
\end{array}\right.\label{EQ.(2.5)}\\
&=&\left\{
\begin{array}{lc}
\rho_n \int_0^L dS \left[\mu_n(k;S)-2\mu_n(k-1;S)+\mu_n(k-2;S)\right],&\qquad k\geq 2 \\
\rho_n \int_0^L dS \left[\mu_n(1;S)-2\mu_n(0;S)+1 \right],&\qquad k=1 \\
1 - \rho_n\int_0^L dS \left[1-\mu_n(1;S) \right],&\qquad k=0 
\end{array}\right.
\label{EQ.(2.6)}
\end{eqnarray}
where Eq.(\ref{EQ.(2.6)}) follows from Eq.(\ref{EQ.(1.11)}), integration by parts and the following limits which result from the existence of the average:
\begin{eqnarray}
&&\lim_{S\to+\infty}S\left[\mu_n(k;S)-2\mu(k-1;S)+\mu_n(k-2;S)\right]=0,\qquad k\geq2\nonumber\\
&&\lim_{S\to+\infty}S\left[\mu_n(1;S)-2\mu(0;S)+1\right]=0,\qquad k=1\label{EQ.(2.7)}\\
&&\lim_{S\to+\infty}S\left[1-\mu_n(0;S)\right]=0,\qquad k=0\nonumber
\end{eqnarray}
Then, $\sigma_n^2(\rho_n L)$ is described in terms of $\mu_n(k;S)$ as
\begin{eqnarray}
\sigma_n^2(\rho_n L)&=&\sum_{k=0}^{+\infty} k^2 e_n(k;\rho_n L)-\rho_n^2  L^2\nonumber\\
&=&\rho_n\int_0^L dS \sum_{k=1}^{+\infty}(2k+1)\left[ \mu_n(k-1;S)-\mu_n(k;S) \right]
+\rho_n\int_0^L\left[1-\mu_n(0;S)\right]dS -\rho_n^2 L^2\nonumber\\
&=&\rho_n L + 2\int_0^L dS \sum_{k=0}^{+\infty}\rho_n\mu_n(k;S) -\rho_n^2 L^2,
\label{EQ.(2.8)}
\end{eqnarray}
where we have used Eq.(\ref{EQ.(1.14)}) and the relation 
\begin{equation}
\sum_{k=0}^{+\infty}\left[(k-1)\mu_n(k-1;S)-k\mu_n(k;S)\right] = 0
\label{EQ.(2.9)}
\end{equation}
with $\mu_n(-1;S)=0$.  Therefore, in the limit $N\to+\infty$, we have the convergence 
\begin{equation}
\Sigma_{\bar{\mu}}^2(L) = 
\lim_{N\to+\infty} \sum_{n=1}^N \sigma_n^2(\rho_n L) = 
L + 2\int_0^L dS \sum_{k=0}^{+\infty}\bar{\mu}(k;S),
\label{EQ.(2.10)}
\end{equation}
where the limit (\ref{EQ.(2.10)}) follows from Assumption (ii) and the following property results from Assumption (i):
\begin{equation}
\sum_{n=1}^N\rho_n^2 \leq \max_n \rho_n
\sum_{n=1}^N\rho_n =\max_n\rho_n\to 0.
\label{EQ.(2.11)}
\end{equation}
\indent
We note the spectral rigidity $\Delta_3(L)$, which was introduced by Dyson and Mehta\cite{DM63}.  In combination with the LNV, this quantity has played a major role in the study of the long-range spectral statistics.  According to Pandey\cite{Pa79} and Eq.(\ref{EQ.(2.10)}), 
the spectral rigidity is described in terms of $\bar{\mu}(k;S)$ as
\begin{eqnarray}
\Delta_{3,\bar{\mu}}(L)
&=&\frac{2}{L^4}\int_0^L dx(L^3-2 L^2 x +x^3)\Sigma^2_{\bar{\mu}}(x)\nonumber\\
&=&\frac{L}{15}+\frac{4}{L^4}\int_0^L dx(L^3-2L^2x+x^3)\int_0^x dS \sum_{k=0}^{+\infty
}\bar{\mu} (k;S)
\label{EQ.(2.12)}.
\end{eqnarray}
\section{Properties of limiting level number variance}
\label{sect3}
We analyze the slope of the limiting LNV
\begin{equation}
g(L) = \frac{1}{L} \Sigma^2_{\bar{\mu}}(L) 
= 1 + \frac{2}{L}\int_0^L dS \sum_{k=0}^{+\infty}\bar{\mu}(k;S),
\label{EQ.(3.2)}
\end{equation}
which has the following convergence: 
\begin{equation}
\lim_{L\to+\infty}g(L)= 1 + 2\sum_{k=0}^{+\infty}\bar{\mu}(k;+\infty)\geq 1+2\bar{\mu}(0;+\infty).
\end{equation}
Since $\mu_n(k;S)$ is monotonically increasing for $S\geq0$, monotonically decreasing for $k=0,1,2,\cdots$, and $0\leq\mu_n(k;S)\leq1$, $\bar{\mu}(k;S)$ has the same properties and is bounded by $\bar{\mu}(0;+\infty)$ as
\begin{equation}
0\leq \bar{\mu}(k;S)\leq \bar{\mu}(0;S)\leq \bar{\mu}(0;+\infty) \leq 1,
\label{EQ.(3.1)}
\end{equation}
where $\bar{\mu}(0;+\infty)$ classifies the cumulative NNLSD (\ref{EQ.(1.2)}) into three cases: Case 1, the Poisson distribution if $\bar{\mu}(0;+\infty)=0$; Case 2, the asymptotic Poisson distribution if $0<\bar{\mu}(0;+\infty)<1$; and Case 3, the sub-Poisson distribution if $\bar{\mu}(0;+\infty)=1$.\\
\indent 
Then, the property of the limiting LNV is evaluated for Cases 1---3 as follows:\\
\indent
Case 1, $\bar{\mu}(0;+\infty)=0$: The limiting LNV $\Sigma_{\bar{\mu}}^2(L)$ agrees with the LNV of Poisson statistics: $\Sigma_{\mbox{\tiny Poisson}}^2(L)=L$.  Note that this condition is equivalent to $\sum_{k=0}^{+\infty}\bar{\mu}(k;S)=0$ since $\bar{\mu}(k;S)$ is monotonically increasing for $S$ and decreasing for $k$.\\
\indent
Case 2, $0<\bar{\mu}(0;+\infty)<1$: $\Sigma_{\bar{\mu}}^2(L)$ possibly deviates from $\Sigma_{\mbox{\tiny Poisson}}^2(L)$ in such a way that the slope of the limiting LNV is $1$ at $L=0$, increases monotonically with $L$, and approaches a number $1+2\bar{\mu}(0;+\infty)$ or more as $L\to+\infty$.\\
\indent
Case 3, $\bar{\mu}(0;+\infty)=1$: $\Sigma_{\bar{\mu}}^2(L)$ possibly deviates from $\Sigma_{\mbox{\tiny Poisson}}^2(L)$ in such a way that the slope of the limiting LNV is $1$ at $L=0$, increases monotonically with $L$, and approaches a number $3$ or more as $L\to+\infty$.\\
\indent
One has Case 1 if the NNLSD of individual components are derived from the scaled distribution functions $\varphi_n(0;S)$ as
\begin{equation}
\mu_n(0;S)=\rho_n\int_0^S \varphi_n(0;\rho_n x)dx,
\label{EQ.(3.5)}
\end{equation}
where $\varphi_n(0;\rho_n S )=p_n(0;S)/\rho_n$ and satisfy
\begin{equation}
\int_0^{+\infty}\varphi_n(0;x)dx=1,\quad\int_0^{+\infty}x \varphi_n(0;x)dx= 1,
\label{EQ.(3.6)}
\end{equation}
and are uniformly bounded by a positive constant $D$ :  $\left| \varphi_n(0;S)\right|\leq D$ ( $1\leq n\leq N$ ).  Indeed, the following holds:
\begin{equation}
\left| \mu(0;S)\right|\leq \sum_{n=1}^N \rho_n^2\int_0^S\left| 
\varphi_n(0;\rho_n x)\right| dx 
\leq DS \sum_{n=1}^N \rho_n^2\leq DS\max_{n}\rho_n
\sum_{n=1}^N\rho_n\to 0\equiv\bar{\mu}(0;S).
\label{EQ.(3.7)}
\end{equation}
Such a bounded condition is possible when the individual spectral components are sparse enough.\\
\indent
In general, one may expect Cases 2 or 3 with $\bar{\mu}(0; S)>0$, which corresponds 
to strong accumulation of energy levels, leading to a singular NNLSD of the individual components.  Such accumulation is expected to arise from the symmetry of the system.  In the next section, we will 
analyze the LNV of the rectangular billiard systems which is known to deviate from Poisson statistics.
\section{Rectangular billiard system}
\label{sect4}
We present our results on the various statistical measures discussed in the 
previous section for a rectangular billiard system whose spectral statistics 
has been precisely analyzed in a number of 
works\cite{BT77,Be85,CK97,CCG85,RV98,MT03,Sh89,BAL91}. 
The eigenenergy levels of this system are given by 
\begin{equation}
\epsilon_{n,m}=n^2+\alpha m^2,\label{EQ.(4.1)}
\end{equation}
where $n$ and $m$ are positive integers, and $\alpha$ is denoted by the lengths of two sides $a$ and $b$ as $\alpha=a^2/b^2$.  The unfolding transformation $\{\epsilon_{n,m}\} \to \{\bar{\epsilon} _{n,m}\}$ is carried out by using the leading Weyl term of the integrated density of states, $\mathcal{N}(\epsilon)$, as 
\begin{equation}
\bar{\epsilon}_{n,m}={\mathcal{N}}(\epsilon_{n,m})
=\frac{\pi}{4\sqrt{\mathstrut\alpha}}\epsilon_{n,m}.\label{EQ.(4.2)}
\end{equation}
Berry and Tabor observed that the NNLSD of this system possibly deviates from Poisson statistics when $\alpha$ is rational\cite{BT77}.  
In this paper, we study irrational cases in addition to a rational case($\alpha=1$) that are described by a finite continued fraction of the 
golden mean $(\sqrt{\mathstrut 5}+1)/2$,
\begin{equation}
\alpha = 1+\frac{1}{1+}\frac{1}{1+}\cdots
\frac{1}{1+}\frac{1}{1+\delta}=\left[1;1,1,\cdots,1,1+\delta\right],
\label{EQ.(4.3)}
\end{equation}
with an irrational truncation parameter $\delta\in [0,1)$.\\
\indent
Fig.1 shows the plots of the LNV $\Sigma^2(L)$ for $\alpha$ corresponding to the (a) 25th, (b) 8th, and (c) 4th 
approximations of the golden mean, and (d) $\alpha=1$.  Our analysis is valid in the 
region $L << L_{\mbox{\tiny max}}$, where 
$L_{\mbox{\tiny max}}=\sqrt{\mathstrut\pi\bar{\epsilon}_{nm}}\alpha^{-1/4}$ 
for the rectangular billiard\cite{Be85,CCG85}.  We used energy levels 
$\bar{\epsilon}_{n,m}\in [4000\times10^7,4001\times10^7]$, which correspond 
to $L_{\mbox{\tiny max}}\sim 3.1\times 10^4$.   The numerical computation was carried out using a 
double precision real number operation.  When the continued fraction is close to the 
golden mean, $\Sigma^2(L)$ is well approximated by 
$\Sigma^2_{\mbox{\tiny Poisson}}(L)=L$ [plot (a)].  On the other hand, in cases in which the 
continued fractions are far from the golden mean, $\Sigma^2(L)$ clearly deviates from 
$\Sigma^2_{\mbox{\tiny Poisson}}(L)$ [plots (b) -- (d)].\\
\indent
Fig.2 shows the slope $g(L)$ of the LNV for the four values of $\alpha$ corresponding to the plots (a)--(d) in Fig.1.  
When $\Sigma^2(L)$ is well approximated by $\Sigma^2_{\mbox{\tiny Poisson}} (L)$, the slope $g(L)$ is $1$ [plot (a)].  Since $\sum_{k=0}^{+\infty}\bar{\mu}(k;S)=0$ is equivalent to $\bar{\mu}(0;S)=0$, this result corresponds to Case 1 given in the previous section(see also Eq.(\ref{EQ.(3.2)})).  In the case in which $\Sigma^2(L)$ deviates from $\Sigma^2_{\mbox{\tiny Poisson}}(L)$, the slope $g(L)$ is $1$ at $L=0$ and increases monotonically with $L$ [plots (b) -- (d)].  In the limit of $L\to+\infty$, plot (b) approaches a number less than $3$ and this result clearly corresponds to Case 2, while plots (c) and (d)  approach numbers greater than $3$ and these results correspond to Case 2 or 3.\\
\indent 
In order to clarify the attribute of plots (c) and (d), we consider
\begin{equation}
\tilde{\mu}(S) = 1 - \frac{1-M(S)}{1-\int_0^S (1-M(x))dx}.\label{EQ.(4.4)}
\end{equation}
that is obtained by the cumulative NNLSD $M(S)=\int_0^S P(0;x)dx$.  The function $\tilde{\mu}(S)$ is equivalent 
to $\bar{\mu}(0;S)$ in the semiclassical limit $\epsilon\to+\infty$. \\
\indent
Figs.3(a)--3(d) show $-\ln{[1-M(S)]}$ for the four values of $\alpha$ corresponding to the plots (a)--(d) in Figs.1 and 2, respectively.  
The dotted line in each figure corresponds to the cumulative Poisson distribution $M_{\mbox{\tiny Poisson}}(S)=1-\exp{(-S)}$.  
Plots (a)--(d) in Fig.4 show $\tilde{\mu}(S)$ for the four values of $\alpha$ corresponding to Figs.3(a)--3(d), respectively.  
When $\Sigma^2(L)$ approximates $\Sigma^2_{\mbox{\tiny Poisson}}(L)$, $M(S)$ fits $M_{\mbox{\tiny Poisson}}(S)$ 
very well[Fig.3(a)].    In this case,  $\tilde{\mu}(S)$ is obviously $0$ [plot (a) in Fig.4].  In case that $\Sigma^2(L)$ deviates 
from $\Sigma^2_{\mbox{\tiny Poisson}}(L)$, $M(S)$ for small value of $S$ clearly deviates from 
$M_{\mbox{\tiny Poisson}}(S)$[Figs.3(b)--3(d)].  However, for large value of $S$, it approaches a line whose 
slope is $1$(see the dashed line in Figs. 3(b)--3(d)).  In these cases, 
$\tilde{\mu}(S)$ in $S\to+\infty$ approaches a number $\tilde{\mu}(+\infty)$ such 
that $0<\tilde{\mu}(+\infty)<1$ [plots (b) -- (d)  in Fig.4].   
Therefore, plots (b)--(d) correspond to Case 2.\\
\indent
For the rectangular billiard in the finite energy  region, we have not yet succeeded in observing Case 3.  
This case is expected to arise in a square billiard ($\alpha=1$) in the high energy limit where a stronger accumulation 
of levels is generated.  Based on the number-theoretical result of Landau (1908)\cite{Landau1908}, 
Connors and Keating have proved that the eigenenergy levels $\epsilon_{n,m}$ of square billiard 
show a logarithmic increase in the mean degeneracy of levels as $\epsilon\to+\infty$,
which is described as
\begin{equation}
1-M(+0) \simeq \frac{4}{\pi}\frac{C_2}{\sqrt{\mathstrut\ln \epsilon}}\rightarrow 0,
\label{EQ.(4.5)}
\end{equation}
where $C_2$ converges to give $C_2\simeq 0.764$\cite{CK97}.  The above limit 
corresponds to a delta function of NNLSD, $P(S)=\delta(S)$, and is consistent with 
$\bar{\mu}(0;S=+0)=\lim_{\epsilon \to +\infty}M(+0)= 1$, which indicates an extremely slow approach 
to Case 3 in $\epsilon\to+\infty$.\\
\indent
Fig.5 shows the cumulative NNLSD $M(S)$ for $\alpha=1$.  This function is not smooth at the level spacings 
separated by a step $\pi/4$\cite{BT77}, which correspond to accumulation of levels.  Note that  
$M(+0)>0$ due to the degeneracy at $S=0$.  Since $\tilde{\mu}(+0)=M(+0)>0$, this 
degeneracy at $S=0$ is identified also in Fig.4.  As the eigenenergy levels become higher, 
$M(+0)$ increases monotonically and approaches $1$.   In the 
limit $\epsilon \to +\infty$ where $M(+0)=\bar{\mu}(0,S=+0)=1$, all steps except the 
step at $S=0$ are suppressed since $M(S)$ is monotonically increasing for $S>0$. \\
\indent
Fig.6 shows $1-M(+0)$ vs $4C_2/\pi\sqrt{\mathstrut\ln \epsilon}$ 
for various energy ranges.  Although we are not yet far enough in the 
high energy region where $1-M(+0)=1-\tilde{\mu}(+0)<<1$, the 
agreement between them is very good, and is better as $\epsilon \to +\infty$.   
Therefore, the extremely slow convergence to Case 3 is well 
reproduced by a numerical computation.  The almost same results for 
$\alpha=22/21$ have already been reported by Robnik and Veble\cite{RV98}.
\section{Conclusion and Discussion}
\label{sect5}
Based on the approach of Berry and Robnik, we have investigated the energy level statistics 
of classically integrable quantum system and discussed its deviations from Poisson statistics.  
In the Berry-Robnik approach, individual eigenstates localizing on the different phase space 
regions provide mutually independent contributions to the statistics of energy levels in 
the semiclassical limit.  Since the phase space of integrable system is densely covered with 
invariant tori, the eigenfunctions of the classically integrable quantum system are localized 
on the regions that have infinitesimal volumes in Liouville measure.  Therefore, we have 
considered the situation in which the eigenenergy sequence is a superposition of infinitely 
many independent components, and each of which gives an infinitesimal contribution to the 
level statistics.  Moreover, by developing the approach of Makino {\it{et al}}\cite{MT03} into the 
statistics of higher order (long-range) spectral fluctuations, the LNV of systems consisting 
of infinitely many spectral components are obtained.  The LNV is characterized by the monotonically 
increasing functions $\bar{\mu}(k;S),k=0,1,2,\cdots$, of the level-spacing $S$, where the lowest 
order term $\bar{\mu}(0;S)$ is associated with the NNLSD.  The property of the LNV is classified 
as follows: Case 1, $\bar{\mu}(0;+\infty)=0$ where the cumulative NNLSD is the Poisson distribution, the LNV 
is the Poissonian $\Sigma^2_{\mbox{\tiny Poisson}}(L)=L$; Case 2, $0<\bar{\mu}(0;+\infty)<1$ 
where the cumulative NNLSD is the asymptotic Poisson distribution, the LNV deviates from 
$\Sigma^2_{\mbox{\tiny Poisson}}(L)$ in such a way that the slope is greater than $1$ and 
approaches a number $\geq 1+2\bar{\mu}(0;+\infty)$ as $L\to+\infty$; Case 3, $\bar{\mu}(0;+\infty)=1$ 
where the cumulative NNLSD is the sub-Poisson distribution, the LNV deviates from $\Sigma^2_{\mbox{\tiny Poisson}}(L)$ in such a way that the slope is greater than $1$ 
and approaches a number $\geq 3$ as $L\to+\infty$.   Therefore, we 
have shown that deviations from Poisson statistics (Cases 2 and 3) are possibly observed, not 
only in the property of NNLSD characterizing short-range spectral fluctuation as shown in 
the work of Makino {\it{et al}}\cite{MT03}, but also in the property of the LNV characterizing 
the fluctuations of all ranges.\\
\indent
Note that Cases 2 and 3 may arise when there is a strong accumulation of levels, which is 
characterized by the singular NNLSD of individual components.  Such accumulation would be 
expected when there is a spatial symmetry or time-reversal symmetry.  One example is a 
rectangular billiard shown in Section \ref{sect4} where the result shows Case 2 in addition 
to Case 1, and an extremely slow approach to Case 3 .  Similar results due to a spatial 
symmetry are also shown in the equilateral-triangular billiard\cite{BW84,Ag08}, 
torus billiard\cite{RB81,RV98}, and integrable Morse oscillator\cite{Sh89}.  Another 
example is a certain type of systems with the time-reversal symmetry studied by 
Shnirelman\cite{Sh75}, Chirikov and Shepelyansky\cite{CS95}, and Frahm and 
Shepelyansky\cite{FS97}.  In this system, the strong accumulation of energy levels, 
resulting from the time reversibility, is reflected in the NNLSD as a sharp 
Shnirelman peak at small level-spacings.\\
\indent
Rigorous results confirming the Berry--Tabor conjecture (Case 1) are also reported for a 
certain classically integrable system.  In the work of Marklof\cite{Ma01} and 
Eskin {\it{et al}}\cite{Es}, eigenvalue problems are reformulated as lattice point 
problems, and it is exactly proved under explicit diophantine conditions that a two-point 
spectral correlation exhibits Poisson statistics.\\
\indent
The periodic-orbit theory, applied for an integrable system with degeneracy in orbit 
actions, gives a similar result.  In the work of Biswas {\it{et al}}\cite{BAL91}, it was 
shown that the slope $g_{\mbox{\tiny{av}}}(L)$ of the LNV $\Sigma^2(L)=g_{\mbox{\tiny{av}}}(L) L$ 
represents the average degeneracy of actions of periodic orbits, which is greater 
than $1$ for the rectangular billiard with a rational ratio $\alpha$ of squared 
sides.  Since this property is qualitatively consistent with the property of the limiting 
LNV (\ref{EQ.(2.10)}) in Cases 2 and 3, one might expect to have the relation:
\begin{equation}
g_{\mbox{\tiny{av}}}(L) = 1 + \frac{2}{L}\int_0^L \sum_{k=0}^{+\infty}\bar{\mu}(k;S)dS,
\label{EQ(5.1)}
\end{equation}
which enables us to discuss the properties of the overall NNLSD from the periodic-orbit 
theory.  From Eqs.(\ref{EQ(5.1)}) and (\ref{EQ.(3.1)}), we expect that the Case 1 
corresponds to the non-degeneracy of actions: $g_{\mbox{\tiny{av}}}(L)=1$, while Cases 2 and 3 
correspond to the degeneracy of actions:$g_{\mbox{\tiny{av}}}(L)>1$.  Therefore, we 
have confirmed that Cases 2 and 3 are closely related to the degeneracy 
in orbit actions.  At the same time, we have only superficially examined 
the possibility that the degeneracy of actions, induced by the 
symmetry, could be an essential factor for the singularity of individual 
spectral components that yields Cases 2 and 3.  This part should be 
investigated in detail in a future work.\\
\indent
There is another approach similar to the one presented in this paper, in which 
the LNV is derived in the limit of infinitely many components\cite{MT05}.  The LNV is described 
by the Dyson two-level cluster function $Y_2$ as
\begin{equation}
\Sigma^2(L) = L -2\int_0^L (L-S) Y_2(S)dS.
\end{equation}
For spectral superposition, $Y_2$ is described in terms of the cluster function of 
individual components $y_{2,n}$ as in Pandey's work\cite{Pa79}
\begin{equation}
Y_2(S)=\sum_{n=1}^N \rho_n^2 y_{2,n}(\rho_n S).
\end{equation}
Then, for the overall LNV, one has the convergence
\begin{eqnarray}
\lim_{N\to+\infty}\Sigma^2(L)= L + 2\int_0^L \bar{c}(S)dS
\label{limit5-1}
\end{eqnarray}
with 
\begin{equation}
\bar{c}(S)=- \lim_{N\to+\infty}\sum_{n=0}^N \rho_n \int_0^{\rho_n S} y_{2,n}(x) dx,
\end{equation}
where $\bar{c}(S)=0$ corresponds to Poisson statistics and $\bar{c}(S)\not=0$ 
indicates deviations from Poisson statistics.  Since $y_{2,n}$ is associated with the 
level-spacing distribution of individual components as $y_{2,n}(x)=1- \sum_{k=0}^{+\infty}p_n(k;x)$, 
$\bar{c}(S)$ is rewritten as 
\begin{equation}
\bar{c}(S)= \lim_{N\to+\infty}\sum_{n=0}^N \rho_n \int_0^{\rho_n S}\sum_{k=0}^{+\infty}p_n(k;x)dx.
\end{equation}
When $p_n(k;S), k=0,1,2,\cdots$, are assumed to satisfy the commutation relation $\int_0^{\rho_n x}\sum_{k=0}^{+\infty}p_n(k;S)dS=\sum_{k=0}^{+\infty}\int_0^{\rho_n x}p_n(k;S)dS$, $\bar{c}(S)$ is described as $\bar{c}(S)=\sum_{k=0}^{+\infty}\bar{\mu}(k;S)$ and limit (\ref{limit5-1}) is consistent with the limit (\ref{EQ.(2.10)}). \\
 \\
Acknowledgment\\
H.M. is grateful to Prof. M. Robnik, Prof. A. Shudo, and Prof. Y. Aizawa for helpful suggestions.  This work is partially supported by KAKENHI(No.18740241 to H.M., No.17540100 to N.M., No.17540365 to S.T.), and by a Grant-in-Aid for the ``Academic Frontier'' Project at Waseda University from the Ministry of Education, Culture, Sports, Science and Technology of Japan.
\appendix
\section{Derivation of Eqs.(\ref{EQ.(2.3)}) and (\ref{EQ.(2.4)})}
\label{appendix}
We briefly show the detailed process to derive Eqs.(\ref{EQ.(2.3)}) and (\ref{EQ.(2.4)}) 
from Eqs.(\ref{EQ.(1.5)}), (\ref{EQ.(1.7)}), and (\ref{EQ.(1.8)}).  First, we 
rewrite Eq.(\ref{EQ.(1.3)}) as 
\begin{equation}
\sum_{k=0}^{+\infty}(k-L)^2E_N(k;L)=\sum_{k=0}^{+\infty}k^2 E_N(k;L) -L^2, 
\label{(A1)}
\end{equation}
where we have used the normalization conditions $\sum_{k=0}^{+\infty}E_N(k;L)=1$ 
and $\sum_{k=0}^{+\infty}k E_N(k;L)=L$, which follow from Eqs.(\ref{EQ.(1.7)}) 
and (\ref{EQ.(1.8)}), and the relations $k=\sum_{n=1}^N k_n$ and $\sum_{n=1}^N\rho_n =1$ as 
\begin{equation}
\sum_{k=0}^{+\infty}E_N(k;L)=\prod_{n=1}^N \sum_{k_n=0}^{+\infty}e_n(k_n;\rho_n L)=1,
\label{(A2)}
\end{equation}
\begin{eqnarray}
\sum_{k=0}^{+\infty}kE_N(k;L) 
&=&\sum_{n=1}^N\sum_{k_n=0}^{+\infty}k_n e_n(k_n;\rho_n L)\prod_{i=1,i\not=n}^N\sum_{k_i=0}^{+\infty}e_i(k_i;\rho_iL)\nonumber\\
&=&\sum_{n=1}^N\rho_n L\cdot 1^{N-1}=L.
\label{(A3)}
\end{eqnarray}
By using Eqs.(\ref{(A2)}) and (\ref{(A3)}), $\sum_{k=0}^{+\infty}k^2 E_N(k;L)$ in the right hand side 
of Eq.(\ref{(A1)}) is calculated in a similar manner:
\begin{eqnarray}
&& \sum_{k=0}^{+\infty}k^2 E_N(k;L) \nonumber\\
&=&\sum_{n=1}^N \sum_{\sum_{i=1,i\not=n}^N k_i=0}^{+\infty}\left(\prod_{i=1,i\not=n}^N e_i 
(k_i;\rho_i L)\right) 
\sum_{k_n=0}^{+\infty}k_n^2e_n(k_n;\rho_n L)\nonumber\\
&&+\sum_{n\not=i}^N\sum_{k_n=0}^{+\infty}\sum_{k_i=0}^{+\infty}k_n e_n(k_n;\rho_n L)k_i e_i 
(k_i;\rho_i L)
\nonumber\\
&=&\sum_{n=1}^N\left(\sum_{k_n=0}^{+\infty}k_n^2 e_n(k_n;\rho_n L)\right)\left(\sum_{k_i=0}^ 
{+\infty}e_{i\not=n}
(k_i;\rho_i L)\right)^{N-1}\label{(2.4)}\\
&&+ \left( \sum_{n=1}^N\sum_{k_n=0}^{+\infty}k_n e_n(k_n;\rho_n L)\right)^2 -\sum_{n=1}^N\left 
(\sum_{k_n=0}^
{+\infty}k_n e_n(k_n;\rho_n L)  \right)^2\nonumber\\
&=&\sum_{n=1}^N \sum_{k_n=0}^{+\infty}k_n^2 e_n(k_n;\rho_n L) + L^2-\sum_{n=1}^N(\rho_n L)^2.
\label{(A5)}
\end{eqnarray}
Since the off-diagonal terms $n\not=i$ vanish in the second equality, we have 
\begin{eqnarray}
\sum_{k=0}^{+\infty}k^2E_N(k;L)-L^2&=&\sum_{n=1}^N\left(\sum_{k_n=0}^{+\infty}k_n^2 e_n(k_n;\rho_n L)-(\rho_n L)^2\right)\nonumber\\
&=&\sum_{n=1}^N\sum_{k_n=0}^{+\infty}(k_n-\rho_n L)^2 e_n(k_n;\rho_n L).
\label{(A6)}
\end{eqnarray}
Therefore, the overall LNV of the system with $N$ independent components is described as 
\begin{equation}
\Sigma_N^2(L)=\sum_{n=1}^N \sigma^2_n(\rho_n L),
\label{(A7)}
\end{equation}
where $\sigma^2_n$ is the LNV of the spectral component:
\begin{equation} 
\sigma^2_n(\rho_n L)
=\sum_{k=0}^{+\infty}(k-\rho_n L)^2 e_n(k;\rho_n L).
\label{(A8)}
\end{equation}
\newpage
\begin{figure}
\centerline{
\includegraphics[width=8 cm]
{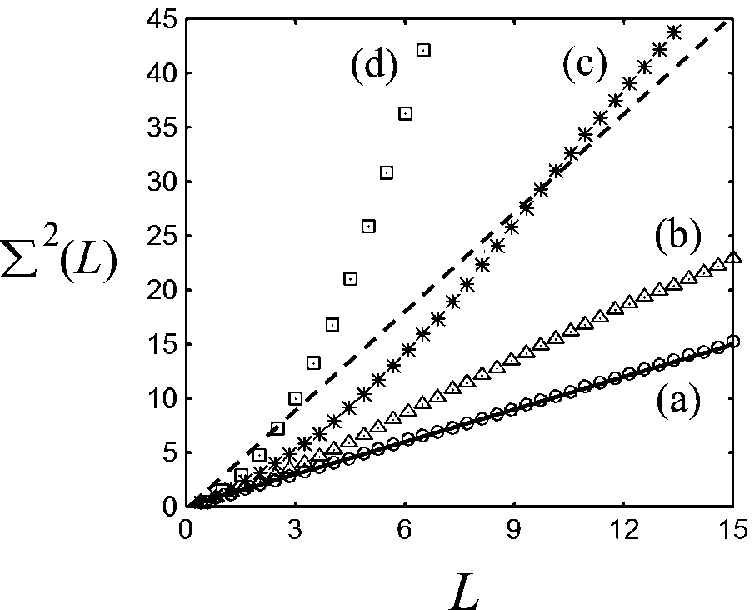}}
FIG.1 Level number variance $\Sigma^2(L)$ of the rectangular billiard systems for 
(a) $25$th, (b) $8$th, (c) $4$th approximations of  $\alpha=(\sqrt{\mathstrut 5}+1)/2$, and (d) $\alpha=1$.  The solid line denotes the LNV of  
Poisson statistics, $\Sigma_{\mbox{\tiny Poisson}}^2(L)=L$.  We used energy levels 
$\bar{\epsilon}_{n,m}\in [4000 \times 10^7,4001 \times 10^7]$.  The total numbers of levels are  
(a) 10000457, (b) 10000428, (c)10000266, and (d) 10000162.  The truncation parameters $\delta$ are 
(a) $\pi\times 10^{-9}$ (b) $\pi/3\times 10^{-7}$ and (c) $\pi\times10^{-9}$.
\label{fig_1}
\end{figure}

\begin{figure}
\centerline{
\includegraphics[width=8 cm]
{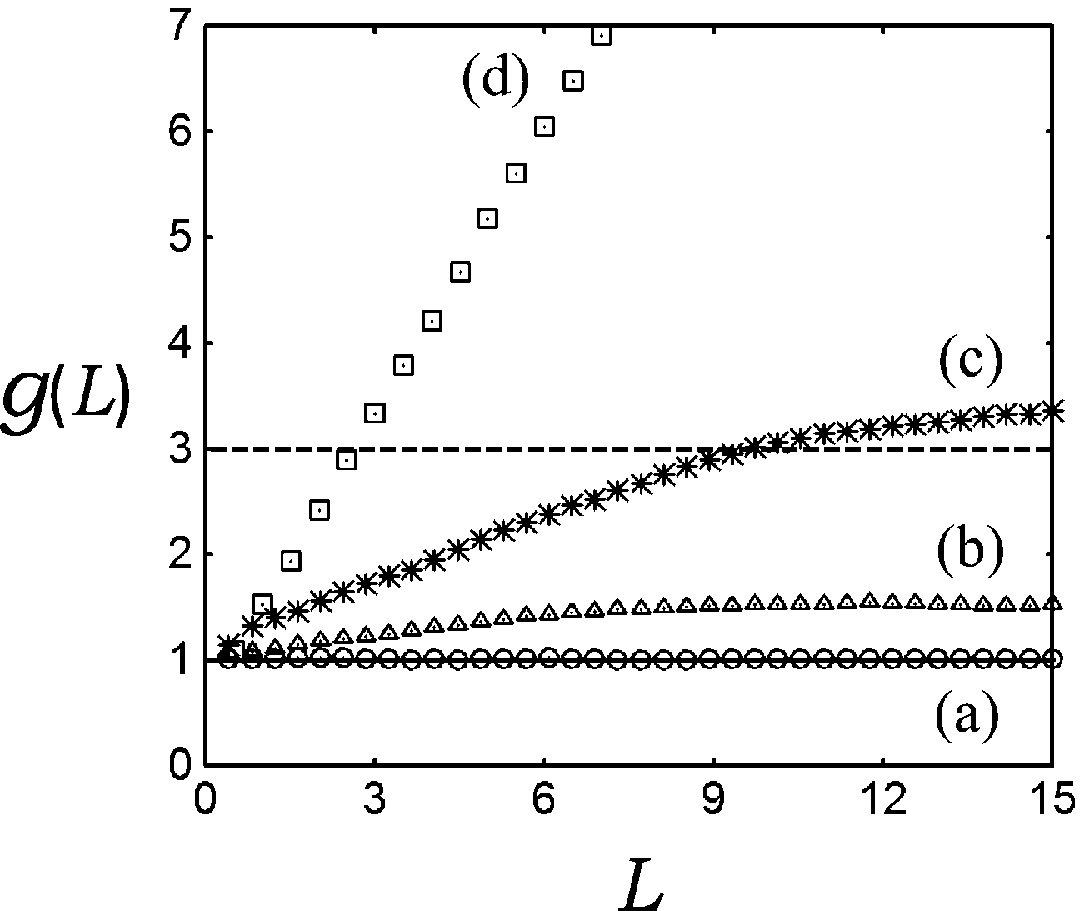}}
{FIG.2 Slope of the level number variance, $g(L)=\Sigma^2(L)/L$, for (a) $25$th, (b) $8$th, (c)  
$4$th approximations of $\alpha=(\sqrt{\mathstrut 5}+1)/2$, and (d) $\alpha=1$.  
The solid line, $g(L)=1$, corresponds to Poisson statistics.}
\label{fig_2}
\end{figure}
\begin{figure}
\centerline{
\includegraphics[width=15 cm]
{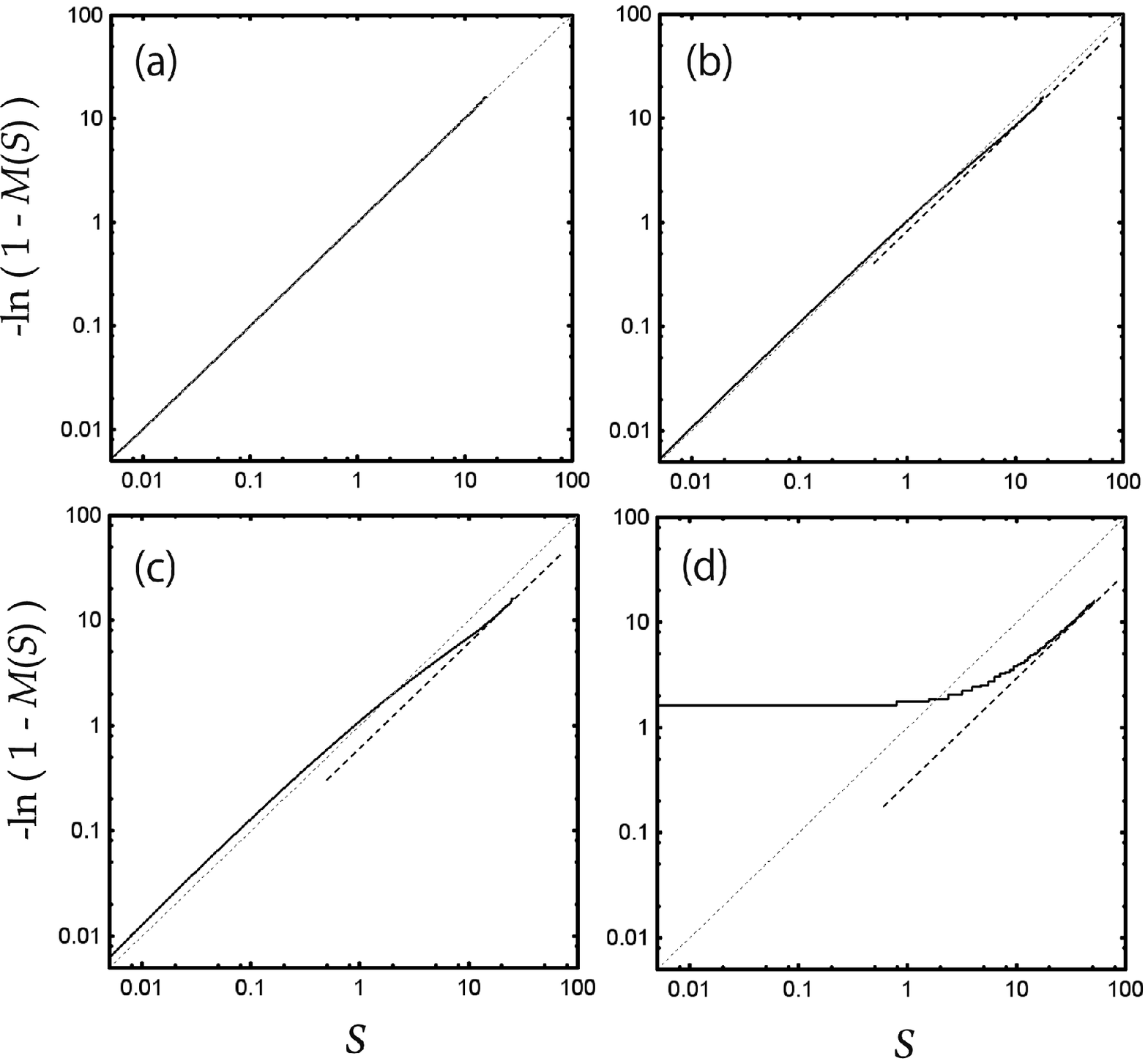}}
{FIG.3 Function $-\ln{[1-M(S)]}$ for (a) $25$th, (b) $8$th, (c) $4$th approximations 
of $\alpha=(\sqrt{\mathstrut 5}+1)/2$, and (d) $\alpha=1$.  The dotted line corresponds to 
the cumulative Poisson distribution $M(S)=1-e^{-S}$.}
\label{fig_3}
\end{figure}
\begin{figure}
\centerline{
\includegraphics[width=8 cm]
{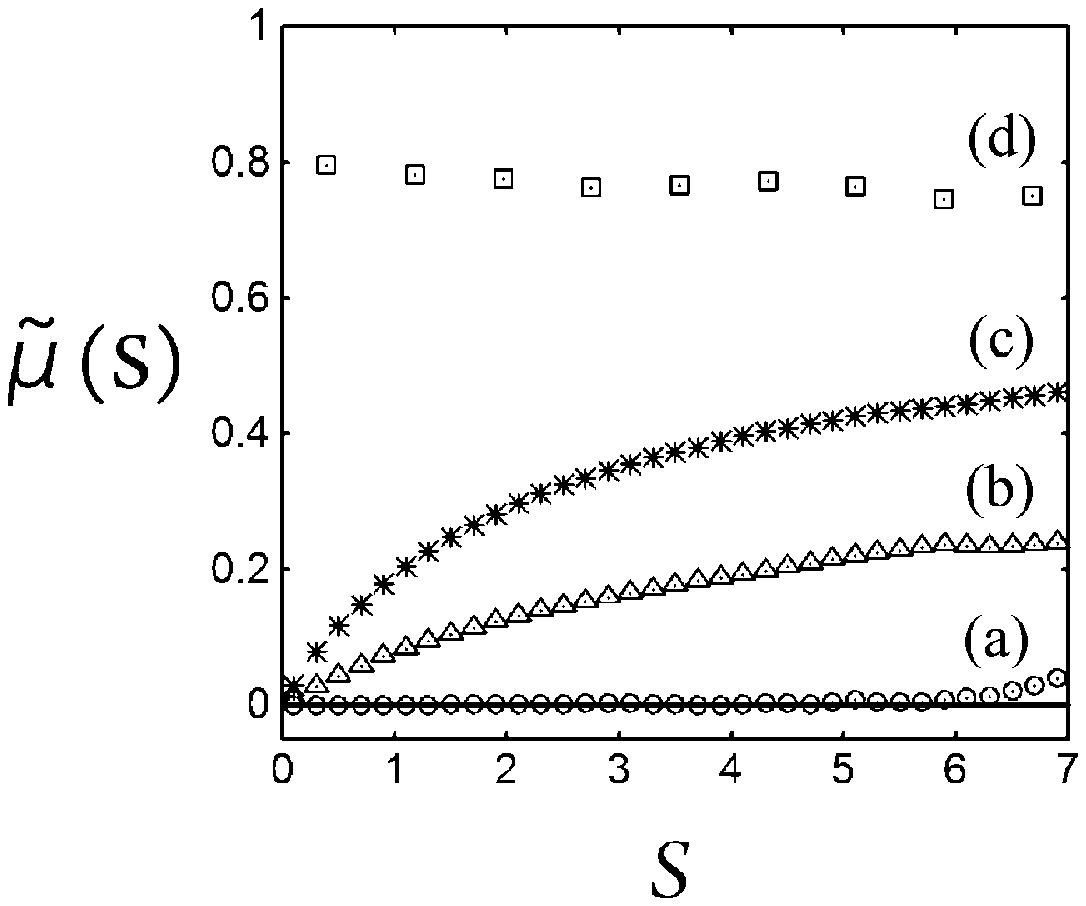}}
{FIG.4 Parameter function $\tilde{\mu}(S)$ for (a) $25$th, (b) $8$th, (c) 
$4$th approximations of $\alpha=(\sqrt{\mathstrut 5}+1)/2$, and (d) $\alpha=1$.  
The solid line, $\tilde{\mu}(S)=0$, corresponds to the Poisson statistics.}
\label{fig_4}
\end{figure}
\begin{figure}
\centerline{
\includegraphics[width=8 cm]
{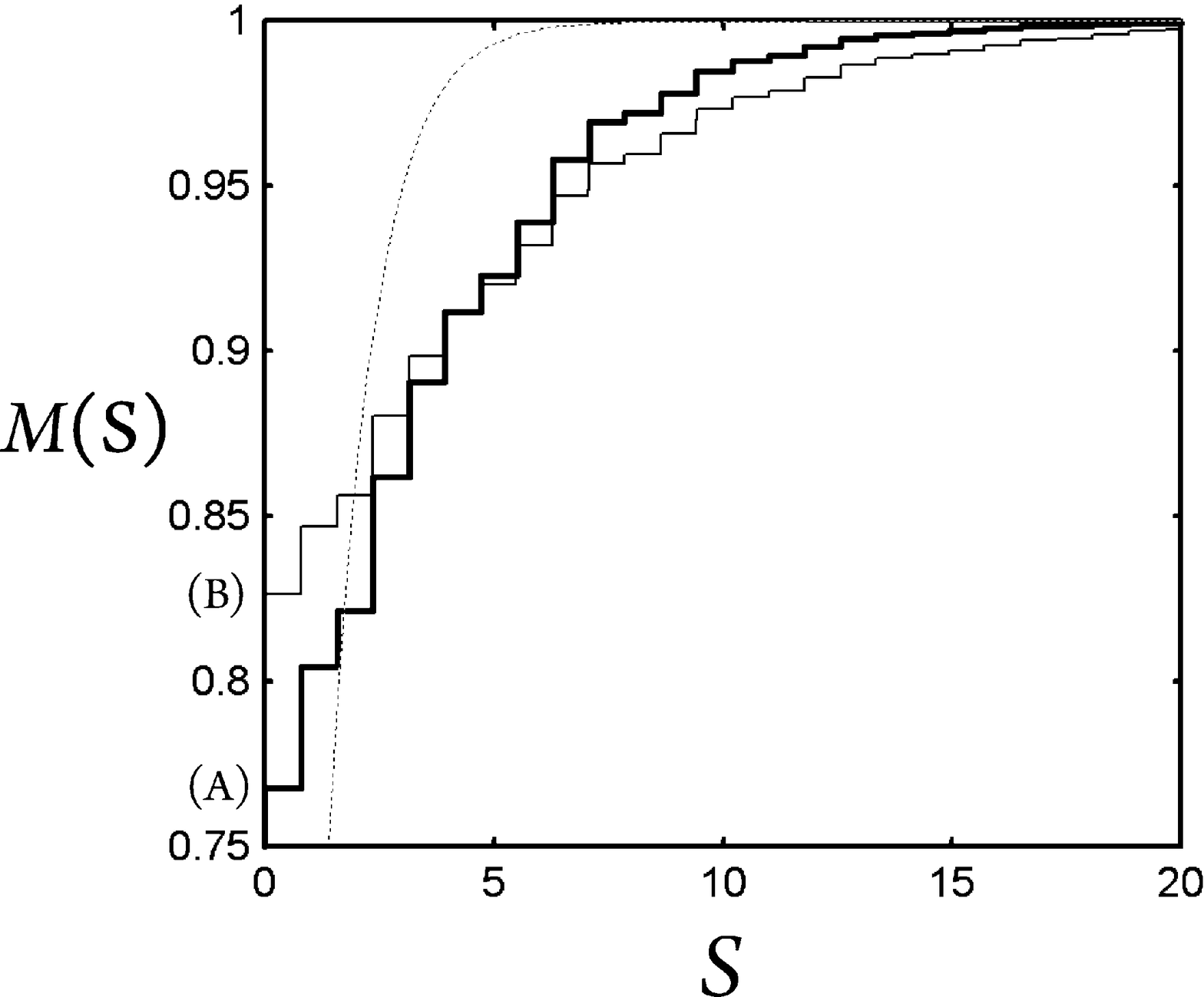}}
{FIG.5 Cumulative NNLSD $M(S)$ for $\alpha=1$.  We used $999844$ 
levels $\bar{\epsilon}_{n,m}\in [40\times 10^6,41\times 10^6]$ for plot (A), $10001872$ levels $\bar{\epsilon}_{n,m}\in [4000000\times 10^7,4000001\times 10^7]$ for plot (B).  We observe 
$M(+0)\simeq 0.767$ for plot (A) and $M(+0)\simeq 0.826$ for plot (B).  The dotted curve is the cumulative Poisson distribution $M(S)=1-e^{-S}$}
\label{fig_5}
\end{figure}
\begin{figure}
\centerline{
\includegraphics[width=8 cm]
{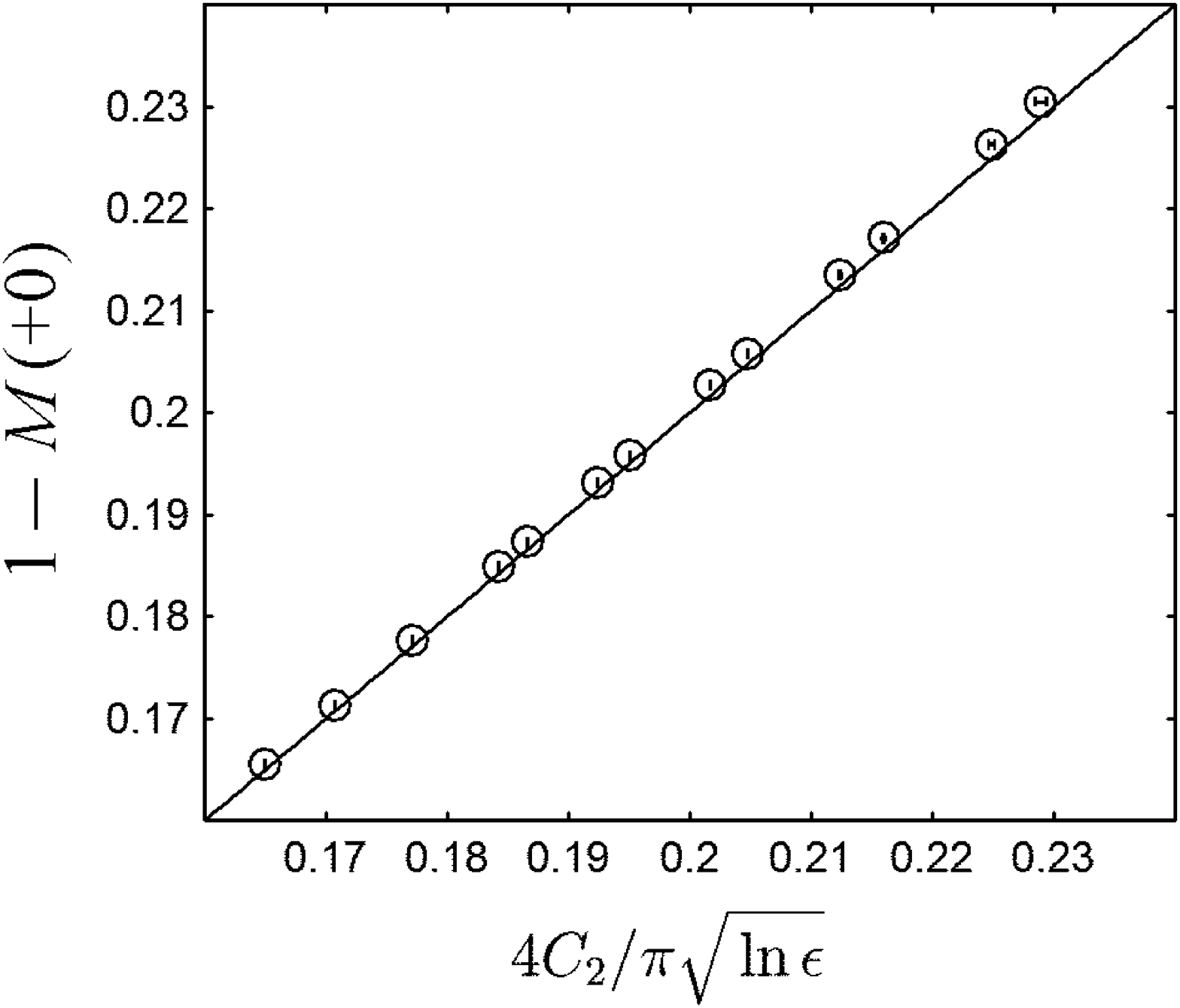}}
{FIG.6 Numerical test of Eq.(\ref{EQ.(4.5)}) for a square billiard ($\alpha=1$).  
The solid line exhibits the theoretical prediction, which is valid in the 
semiclassical limit $\epsilon\to+\infty$.  In each plot, we used $10^7$ eigenenergy levels 
obtained from the unfolded energy range.  The error bar exhibits the energy range 
acquiring numerical data.}
\label{fig_6}
\end{figure}

\begin{thebibliography}{99}
\bibitem{BT77}
M.V.~Berry and M.~Tabor,~Proc.~R.~Soc.~Lond. A~{\bf356},~375~(1977).
\bibitem{BGS84}
O.~Bohigas, M.J.~Giannoni, and C.~Schmit,~Phys.~Rev.~Lett.~{\bf 52},~1~(1984).
\bibitem{ME91}
M.L.~Mehta,{\it{Random Matrices}}(2nd ed., San Diego, CA: Academic Press, 1991)
\bibitem{Bo89}
O.~Bohigas,~{\it{Random matrices and chaotic dynamics}}(LesHouches 1989 Session LII, chaos and quantum physics, North-Holland).
\bibitem{Be85}
M.~V.~Berry,~Proc.~Roy.~Soc.~Lond. A~{\bf 400},~229~(1985).
\bibitem{HO84}
J.~H.~Hannay and A.~M.~Ozorio de Almedia,~J.~Phys.~A~{\bf 17},~3429~(1984).
\bibitem{BT76}
M.V.~Berry and M.~Tabor,~Proc.~R.~Sco.~{\bf 349},~101~(1976);
~M.V.~Berry and M.~Tabor,~J.~Phys.~A {\bf 10},~371~(1977).
\bibitem{Mo81}
S.~A.~Molchanov,~Commun.~Math.~Phys.~{\bf 78},~429~(1981);~S.~A.~Molchanov,~Math.~USSR Izvestija ~{\bf 12},~69~(1978).
\bibitem{Bl90}
P.~M.~Bleher,~J.~Stat.~Phys.~{\bf 61},~869~(1990);~P.~M.~Bleher,~J.~Stat.~Phys.~{\bf 63},~261~(1991).
\bibitem{Sa97}
P.Sarnak, {\it{Values at integers of binary quadratic forms, Harmonic Analysis and Number Theory}} (Montreal, PQ, 1996),~CMS Conf.Proc.{\bf 21}(Amer.~Math.~Soc.,~Providence, RI.),~181~(1997).
\bibitem{CK97}
R.~D.~Connors and J.~P.~Keating,~J.~Phys.~A~{\bf 30},~1817~(1997).
\bibitem{Ma98}
J.~Marklof, ~Comm.~Math.~Phys.~{\bf{199}},~169~(1998).
\bibitem{Ma01}
J.~Marklof, ~Progr. Math.{\bf 202}, Birkhauser,Basel,~421~(2001); DUKE Math. J.~{\bf 115},~409~(2002); Ann. of Math.~{\bf 158},~419~(2003).
\bibitem{Es}
A.~Eskin,~G.A.~Margulis and S.~Mozes, {\it{Quadratic forms of signature (2,2) and eigenvalue spacings on rectangular 2-tori}}, to appear in Ann. Math.
\bibitem{CCG85}
G.~Casati, B.~V.~Chirikov and I.~Guarneri,~Phys.~Rev.~Lett.~{\bf 54},~1350~(1985).
\bibitem{Fe85}
M.~Feingold,~Phys.~Rev.~Lett.{\bf 55},~2626~(1985)
\bibitem{SV86}
T.~H.~Seligman and J.~J.~M.~Verbaarschot, Phys.~Rev.~Lett.~{\bf 56},~2767~(1986).
\bibitem{RV98}
M.~Robnik and G.~Veble,~J.~Phys.~A~{\bf 31},~4669~(1998).
\bibitem{MT03}
H.~Makino and S.~Tasaki,~Phys.~Rev.~E ~{\bf 67},~066205~(2003).
\bibitem{BR84}
M.V.~Berry and M.~Robnik,~J.~Phys.~A {\bf 17},2413 (1984).
\bibitem{Be77}
M.~V.~Berry,J.~Phys.~A~{\bf 10},~2083~(1977).
\bibitem{Ro98}
M.~Robnik,~Nonlinear Phenomena in Complex Systems {\bf 1},~n1,~1~(1998).
\bibitem{Be1977}
M.~V.~Berry,~Phil.~Trans.~R.~Soc.~A~{\bf 287},~237~(1977).
\bibitem{Fe57}
W.~Feller, {\it{An introduction to probability theory and its applications}} (2nd ed., John Wiley \& Sons, Inc., New York, 1957)
\bibitem{LR94}
B.~Li and M.~Robnik,~J.~Phys.~A~{\bf 27},~5509~(1994).
\bibitem{PR93}
T.~Prosen and M.~Robnik, J.~Phys.~A {\bf 26},~2371~(1993); T.~Prosen and M.~Robnik, J.~Phys.~A {\bf 27},~8059~(1994).
\bibitem{Pr96}
T.~Prosen,~Physica~D~{\bf 91},~244~(1996).
\bibitem{PR97}
M.~Robnik and T.~Prosen, J.~Phys.~A {\bf 30},~8787~(1997).
\bibitem{RB81}
P.~J.~Richens and M.~V.~Berry, Physica~(Amsterdam)~2D,~495~(1981).
\bibitem{BW84}
M.~V.~Berry and M.~Wilkinson,~Proc.~R.~Soc.~London A~{\bf 392},~15~(1984).
\bibitem{Sh89}
A.~Shudo,~Prog.~Theor.~Phys.~Suppl.{\bf 98},~173~(1989).
\bibitem{BAL91}
D.~Biswas, M.~Azam, and S.V.~Lawande,~Phys.~Rev.~A~{\bf 43},~5694~(1991);
D.~Biswas, M.~Azam, and S.V.~Lawande,~J.~Phys. A~{\bf 24},~1825~(1991).
\bibitem{Sh75}
A.~I.~Shnirelman,~Usp.~Mat.~Nauk.{\bf 30},~265~(1975).
\bibitem{CS95}
B.~V.~Chirikov and D.~L.~Shepelyansky,~Phys.~Rev.~Lett.~{\bf 74},~518~(1995).
\bibitem{FS97}
K.~M.~Frahm and D.~L.~Shepelyansky, ~Phys.~Rev.~Lett.~{\bf{78}},~1440~(1997).
\bibitem{Gu70}
M.~C.~Gutzwiller,~J.~Math.~Phys.{\bf 11},~1791~(1970);~M.~C.~Gutzwiller,~J.~Math.~Phys.{\bf 12},343 (1971);~M.~C.~Gutzwiller,~{\it Chaos in Classical and Quantum Mechanics}(Springer-Verlag, Berlin, 1990)
\bibitem{MC72}
M.~L.~Mehta and J.~des Cloiseaux, Indian J. Pure Appl. Math.~{\bf 3},~329~(1972).
\bibitem{ABS97}
R.~Aurich, A.~Backer and F.~Steiner,~Int. J. Mod. Phys. B~{\bf 11},~805~(1997).
\bibitem{MI07}
N.~Minami,~CRM Proceeding and Lecture Notes~{\bf 42},~353~(2007).
\bibitem{Po60}
Seminal article by Rozenzweig and Porter of 1960(see also Ref.\cite{ME91})
\bibitem{DM63}
F.~J.~Dyson and M.~L.~Mehta,~J.~Math.~Phys.~{\bf 4},~701~(1963).
\bibitem{Pa79}
A.~Pandey,~Ann.~Phys.~(N.Y.)~{\bf 119},~170~(1979).
\bibitem{Landau1908}
E.~Landau,~ Archiv der Math. und Physik III~{\bf 13},~305~(1908).
\bibitem{Ag08}
F.~M.~de Aguiar,~Phys.~Rev. E~{\bf 77},036201~(2008).
\bibitem{MT05}
H.~Makino and S.~Tasaki,~Prog.~Theor.~Phys.~{\bf 5},~929~(2005).
\end{thebibliography}
\end{document}